\documentclass[%
 reprint,
 amsmath,amssymb,
 aps,
]{revtex4-1}

\usepackage{graphicx}
\usepackage{dcolumn}
\usepackage{bm}
\usepackage{hyperref}
\usepackage[mathlines]{lineno}
\usepackage[usenames]{color}
\usepackage{colortbl}

\usepackage{epstopdf}

\begin{document}

\preprint{APS/123-QED}

\title{Tomographic causal analysis of two-qubit states and tomographic discord}
\author{Evgeny Kiktenko$^{1,2}$}
\author{Aleksey Fedorov$^{1,3}$}
\affiliation{$^{1}$Bauman Moscow State Technical University, 2nd Baumanskaya St. 5, Moscow 105005, Russia}
\affiliation{$^{2}$Geoelectromagnetic Research Center of Schmidt Institute of Physics of the Earth, Russian Academy of Sciences, PO Box 30, Troitsk, Moscow Region 142190, Russia}
\affiliation{$^{3}$Russian Quantum Center, Novaya St. 100, Skolkovo, Moscow 143025, Russia.}

\date{\today}

\begin{abstract}
We study a behavior of two-qubit states subject to tomographic measurement.
In this Letter we propose a novel approach to definition of asymmetry in quantum bipartite state based on its tomographic Shannon entropies.
We consider two types of measurement bases:
the first is one that diagonalizes density matrices of subsystems and is used in a definition of tomographic discord,
and the second is one that maximizes Shannon mutual information and relates to symmetrical form quantum discord.
We show how these approaches relate to each other and then implement them to the different classes of two-qubit states.
Consequently, new subclasses of $X$-states are revealed.
\begin{description}
\item[PACS numbers]
03.65.Wj, 03.65.Ta, 03.67.-a
\end{description}
\end{abstract}

\maketitle

\section{Introduction}\label{sec:Intro}
In general, measurements irreversibly change a state of quantum system.
Quantum tomography is an experimental method, which restores a complete information about an unknown quantum state using preparation of set of its copies and measurements statistics obtained in different bases.

The main feature of quantum tomography is a complete characterization of quantum states and processes {\it directly} from experimental data.
Quantum states of light were completely characterized via the method of balanced homodyne detection (BHD) \cite{Beck}.
These works inspired series of new experiments \cite{Lvovsky3} as well as intensive theoretical work on analysis and improvement of the BHD setup \cite{Lvovsky3}-\cite{DAriano}.
Moreover, quantum tomography was used for characterization of quantum states of current (voltage) in the Josephson junction \cite{MankoOV}.

On the other hand, quantum tomography is an original picture of quantum mechanics, where quantum states are described in terms of nonnegative probability distributions functions \cite{Manko}-\cite{Fedorov}.
Quantum tomography is equivalent to other approaches to quantum mechanics, and tomograms are directly related to quasi-probability distribution functions \cite{Review}-\cite{Review2}.

One of the areas, where quantum tomography is of interest, is consideration of correlation properties in bipartite states.
As a results of purely probabilistic description of states, tomographic version of the Shannon entropy \cite{Shannon} and the R\'enyi \cite{Renyi}-\cite{Renyi2} entropy naturally appear.
Being a bridge between classical information theory and quantum information theory \cite{OVVI}, it allows to use some well-known inequalities for Shannon and R\'enyi entropies for investigation of novel properties of quantum states \cite{Shannon}-\cite{Properties}.
Recently, tomographic approach to quantum discord was suggested \cite{Manko3}.
In particular, tomographic discord for two-qubit $X$-states was considered.
This analysis posed an important problem of relation between original~\cite{Discord} and tomographic discords.

Another interesting question, posed in Ref.~\cite{Zyc}, is about the role of asymmetry between parties of bipartite state in respect to its properties.
Due to such asymmetry, decoherence acting on different parties leads to different rates of correlation decay, so the question about robustness of parties appears.
For the purpose of asymmetry investigation the method of quantum causal analysis was proposed~\cite{Kikt}.
It was successfully implemented to two-\cite{Kikt2} and three-\cite{Kikt3} qubit states and atom-field interaction \cite{Kikt4}, where interesting conclusions were made.

In the current Letter we combine quantum causal analysis with quantum tomography.
We obtain two novel measures of bipartite state asymmetry, based on tomographic discord and symmetric version~\cite{SymDisc} of quantum discord.
We show that tomographic discord is not greater than symmetric quantum discord.
For a demonstration of obtained results we consider the simplest case of bipartite system, and show that even for them nontrivial phenomena occur.

The Letter is organized as follows.
We start from brief consideration of quantum causal analysis in section~\ref{sec:2} and quantum tomography in section~\ref{sec:3}.
In section \ref{sec:4} we show how quantum causal analysis can be modified via tomography.
In section \ref{sec:5} we implement tomographic causal analysis different classes of two-qubit states.
The results of the Letter are summed up in Section\,\ref{sec:6}.

\section{Quantum causal analysis}\label{sec:2}

Quantum causal analysis \cite{Kikt}-\cite{Kikt4} is a formal method for a treatment of informational asymmetry between parties of bipartite states.
The term ``causal'' comes from classical causal analysis (see, e.g.~\cite{CCA}), where a such asymmetry can be related to real causal connection between two processes.
In the quantum domain the conception of causality usually is considered in framework of channels \cite{CausCh} or probability wave propagation \cite{QCaus}.
Nevertheless, it is convenient to introduce formal definitions of ``causes'' and ``effects'' in bipartite states, however, one should understand them only as labels \cite{disc}.

The idea of quantum causal analysis is the following.
Consider a bipartite quantum system $AB$ in the Hilbert space $\mathcal{H}=\mathcal{H}_{A}\otimes\mathcal{H}_{B}$.
It is described via density operator
$\hat{\rho}_{AB}\in\Omega(\mathcal{H}_{AB})$ with $\hat{\rho}_{A}=\mathrm{Tr}_{B}\hat{\rho}_{AB}\in\Omega(\mathcal{H}_{A})$ and $\hat{\rho}_{B}=\mathrm{Tr}_{A}\hat{\rho}_{AB}\in\Omega(\mathcal{H}_{B})$
being corresponding density operators of subsystems.
Here $\Omega(\mathcal{H})$ is the set of positive operators of unit trace (density operators) in a Hilbert space $\mathcal{H}$.
The basic quantity of quantum information theory is the von Neumann entropy given by
$$
	S_{X}\equiv S\left[\hat{\rho}_{X}\right]=-\mathrm{Tr}\left[\hat{\rho}_{X}\log\hat{\rho}_{X}\right], \quad X\in\{A,B,AB\}.
$$
Here we restrict our consideration to finite dimensional Hilbert spaces and take the logarithm to base 2 ({\it i.e.}, we measure entropy in bits).

The amount of correlations between $A$ and $B$ is given by the (symmetric) quantum mutual information
\begin{equation}\label{eq:QMutInf}
	I_{AB}=S_A+S_B-S_{AB}, \quad I_{AB}=I_{BA}.
\end{equation}

To describe a possible asymmetry of correlations we introduce a pair of independence functions
\begin{eqnarray}\label{eq:QIndFunc}
	i_{A|B}=\frac{S_{AB}-S_B}{S_A}=1-\frac{I_{AB}}{S_A}, \nonumber \\
	i_{B|A}=\frac{S_{AB}-S_A}{S_B}=1-\frac{I_{AB}}{S_B}. \nonumber
\end{eqnarray}
which have the following properties:
(i) they take values from $-1$ to $1$ and the less $i_{Y|X}$ is, the stronger $X$ defines $Y$
($i_{Y|X}=-1$ corresponds to maximal quantum correlations, $i_{Y|X}=0$ corresponds to $Y$ being a classical one-valued function of $X$, and $i_{Y|X}=1$ corresponds to $Y$ being independent from $X$);
(ii) negative values correspond to negative conditional entropy and imply a presence of entanglement between partitions;
(iii) for all pure entangled states $\hat{\rho}_{AB}=|\Psi\rangle_{AB}\langle\Psi|$ the both independence functions take minimal values ($i_{A|B}=i_{B|A}=-1$);
(iv) in general, for mixed states relation $i_{A|B}\neq i_{B|A}$ holds.

Further, we can introduce the following formal definitions: in bipartite state $\hat{\rho}_{AB}$ with $S_A\neq S_B$ the party $A$ is the ``cause'' and $B$ is the ``effect'' if $i_{A|B}>i_{B|A}$.
Vice versa, one has $B$ being the ``cause'' and $A$ being the ``effect'' if $i_{B|A}>i_{A|B}$.

Finally, we need to introduce a measure of asymmetry based on independence functions.
In the current Letter it is convenient to use difference
\begin{equation}\label{eq:Qdelta}
	d_{AB}=i_{A|B}-i_{B|A}=I_{AB}\frac{S_A-S_B}{S_A S_B}, \,\, d_{AB}\in(-2,2).
\end{equation}
The zero value is obtained for symmetric or non-correlated states, while the extreme values are obtained in cases when the entropy of one subsystem tends to zero, while the entropy of another does not, and mutual information takes the maximal possible value which is doubled entropy of the first subsystem.

\section{Quantum tomography}\label{sec:3}

Quantum tomography suggests physical picture of quantum mechanics as well as it has interesting mathematical structure.
Mathematical aspects of quantum tomography are well understood in terms of group theory \cite{D'Ariano2}, $C^{*}$ algebra \cite{MankoC} and groupoids \cite{MankoGr}.

Following \cite{D'Ariano2}, we define quantum tomograms thought mapping of $\hat\rho\in\Omega(\mathcal{H})$ on a parametric set of probability distribution functions
\begin{equation}\label{map}
	\hat\rho\in{\Omega(\mathcal{H})}\xrightarrow{\text{G}(g)}\mathcal{T}\{g,m\},
\end{equation}
where $m$ is a physical observable, ${\rm G}(g)$ is a transformation group
with parametrization by $g$,
and parametric set $\mathcal{T}\{g,m\}$ is called quantum tomogram of the state $\hat\rho$.
From physical point of view, every element of parametric set  $\mathcal{T}\{g,m\}$ is a probability of observation value $m$ after transformation ${\rm G}$.

In case of continuous variables, {\it i.e.} when $\dim\mathcal{H}=\infty$, group $\rm{Sp}(2n,\mathbb{R})$ of phase space symplectic transformation plays role of transformation group ${\rm G}(g)$.
Mapping (\ref{map}) at that rate reads
$$
	\mathcal{T}(Q, \mu, \eta)=\langle{Q, \mu, \eta}|{\hat\rho}|{Q, \mu, \eta}\rangle, \quad \hat\rho\in\Omega(\mathcal{H}),
$$
where $|{Q, \mu, \eta}\rangle$ is an eigenvector of the Hermitian operator $\mu\hat{q}+\eta\hat{p}$ for the eigenvalue $Q$.
One can see that $\mathcal{T}(Q, \mu, \eta)$ is positive and normalized on $Q$.
This representation is directly related with star-product quantization \cite{Review} and the Weyl--Heisenberg group \cite{D'Ariano2}.

The BDH setup reduces to mixing on beam splitter of measurable (weak) field and strong coherent field with changing phase $\theta$.
In terms of (\ref{map}) the observable is $\hat{Q}=\widehat{q}\cos\theta+\widehat{p}\sin\theta$,
where angle $\theta\in\mathbb{R}/2\pi\mathbb{Z}$ could be interpreted as rotation angle of the phase space.

In case of system with discrete variables ($\dim\mathcal{H}<\infty$) mapping (\ref{map}) transforms to the following relation
\begin{equation}\label{defd}
	\mathcal{T}_m(U)=\langle{m}|U\hat\rho{U}^{\dagger}|{m}\rangle, \quad \hat\rho\in\Omega(\mathcal{H}),
\end{equation}
where normalization and positivity follows directly from definition (\ref{defd})
$$
	\sum\nolimits_m{\mathcal{T}_m(U)}=1, \qquad {\mathcal{T}_m(U)}\geq 0.
$$

In case $U\in{\rm SU(2)}$ definition (\ref{defd})  reduces to general definition for spin tomograms
$$
	U=\begin{pmatrix}
	\alpha & \beta \\
	-\beta^{*} & \alpha^{*} \\
	\end{pmatrix},
	\quad |\alpha|^2+|\beta|^2=1.
$$

Here $\alpha,\beta\in\mathbb{C}$ are the Cayley--Klein parameters.
In $U\in{\rm SU(2)}$ case the Euler angles \cite{Manko2} and quaternions \cite{Fedorov} can be used for representation of tomograms \cite{Manko4}.

\section{Tomographic approach to quantum causal analysis}\label{sec:4}

Here we suggest to use an approach of quantum causal analysis to bipartite system asymmetry with respect to observable outcomes described by quantum tomography.
The bipartite state tomogram reads
$$
	\mathcal{T}_{AB}(U_A\otimes U_B)=\{\mathcal{T}_{AB_{ij}}(U_A\otimes U_B)\},
$$
and reduced tomograms have the form
\begin{eqnarray}
	\mathcal{T}_{A}(U_A)=\left\{\sum\nolimits_j\mathcal{T}_{AB_{ij}}(U_A\otimes U_B)\right\}, \nonumber \\
	\mathcal{T}_{B}(U_B)=\left\{\sum\nolimits_i\mathcal{T}_{AB_{ij}}(U_A\otimes U_B)\right\}. \nonumber
\end{eqnarray}
Here we imply that two parties are measured in local bases described by $U_A$ and $U_B$.
Since tomograms being classical probability distributions one can introduce the Shannon entropy \cite{Shannon}
$$
		H_X(U)=-\sum\nolimits_m {\mathcal{T}_{X_{m}}(U) \log{\mathcal{T}_{X_{m}}(U)}},
$$
which depends on unitary transformation $U$ or, in other words, on the way we measure the system.

The value of observed correlations in bipartite system $AB$ is described by classical mutual information
\begin{equation}\label{eq:CMutInf}
	J_{AB}(U_A,U_B)=H_A(U_A)+H_B(U_B)-H_{AB}(U_A\otimes U_B)
\end{equation}
being a straightforward analog of~(\ref{eq:QMutInf}).

We see that correlation also depends on the way we ``look'' at bipartite system and the question of the particular measurement setup (particular $U_A$ and $U_B$) appears.
This question is directly relates to the quantum discord which is the difference between quantum mutual information~(\ref{eq:QMutInf}) and observable correlations obtained in spirit of~(\ref{eq:CMutInf}).
Here we can point out two approaches.

\subsection{``Tomographic'' scheme}
First approach is to choose rotation operators in such a way that density matrices of subsystems after rotation become diagonal.
We denote these operators as $U_A^0$ and $U_B^0$ and the corresponding tomogram as $\mathcal{T}_{AB}^\mathrm{tom}=\mathcal{T}_{AB}(U_A^0\otimes U_B^0)$.
The direct corollary of this choice is that subsystem Shannon entropies become equal to their von Neumann analogues
$$
	H_A(U_A^0)=S_A, \quad H_B(U_B^0)=S_B.
$$

This approach corresponds to the {\it tomographic discord} introduced in \cite{Manko3}:
$$
	D^\mathrm{tom}_{AB}=I_{AB}-J_{AB}^\mathrm{tom}=H_{AB}(U_A^0\otimes U_B^0)-S_{AB},
$$
with $J_{AB}^\mathrm{tom}=J_{AB}(U_A^0,U_B^0)$.

The asymmetry is given by following values of ``tomographic'' independence functions
\begin{eqnarray}
	 i_{A|B}^\mathrm{tom}=\frac{H_{A}(U_A^0)-J^\mathrm{tom}_{AB}}{H_{A}(U_A^0)}=1-\frac{J^\mathrm{tom}_{AB}}{S_A}\geq0, \nonumber \\
	 i_{B|A}^\mathrm{tom}=\frac{H_{B}(U_B^0)-J^\mathrm{tom}_{AB}}{H_{B}(U_B^0)}=1-\frac{J^\mathrm{tom}_{AB}}{S_B}\geq0. \nonumber
\end{eqnarray}
We obtain that they are greater of equal to their quantum analogies
$$
	i_{A|B}^\mathrm{tom}-i_{A|B}=\frac{D_{AB}^\mathrm{tom}}{S_A}\geq0, \quad
	i_{B|A}^\mathrm{tom}-i_{B|A}=\frac{D_{AB}^\mathrm{tom}}{S_B}\geq0.
$$
The ``tomographic'' asymmetry defined by
$$
	 d_{AB}^{\mathrm{tom}}=i_{A|B}^\mathrm{tom}-i_{B|A}^\mathrm{tom}=J^\mathrm{tom}_{AB}\frac{S_A-S_B}{S_AS_B}
$$
has the same sign as its quantum analog $d_{AB}$ but with less or equal magnitude
\begin{equation}\label{eq:TomDeltaVSDelta}
	 \frac{d_{AB}^{\mathrm{tom}}}{d_{AB}}=\frac{J_{AB}^\mathrm{tom}}{I_{AB}}=1-\frac{D_{AB}^\mathrm{tom}}{I_{AB}}\leq1.
\end{equation}

Finally, one can see that the conclusions about the asymmetry made with quantum causal analysis using original density operator $\hat{\rho}_{AB}$ and classical causal analysis using tomogram $\mathcal{T}_{AB}(U_A^0\otimes U_B^0)$ qualitatively are the same, however, the measurement process smoothes the original asymmetry on the value of $D_{AB}^\mathrm{tom}$.

\subsection{``Optimal'' scheme}
The second approach is to use bases that {\it optimize} ({\it i.e.}, maximize) the classical mutual information
\begin{eqnarray}
	\{U_A^\mathrm{opt},U_B^\mathrm{opt}\}=\mathrm{arg}\max_{U_A,U_B}{\{J_{AB}(U_A,U_B)\}}.
\end{eqnarray}
We denote the corresponding tomogram as
$$
	\mathcal{T}_{AB}^\mathrm{opt}=\mathcal{T}_{AB}(U_A^\mathrm{opt}\otimes U_B^\mathrm{opt}).
$$

This approach corresponds to the symmetric version of quantum discord, studied in detail in~\cite{SymDisc}
\begin{eqnarray}
	D^\mathrm{opt}_{AB}=I_{AB}-J^\mathrm{opt}_{AB}, \nonumber \\
	J^\mathrm{opt}_{AB}=\max_{U_A,U_B}{J_{AB}(U_A,U_B)} \nonumber
\end{eqnarray}

Analogically, ``optimal'' pair of independence functions has the following form
\begin{equation}
	 i_{A|B}^\mathrm{opt}=1-\frac{J^\mathrm{opt}_{AB}}{H_A^\mathrm{opt}}\geq0, \quad  i_{B|A}^\mathrm{opt}=1-\frac{J^\mathrm{opt}_{AB}}{H_B^\mathrm{opt}}\geq0 \nonumber
\end{equation}
where we use definitions
$$
	H_{A}^\mathrm{opt}=H_A\left(U_A^\mathrm{opt}\right), \quad H_{B}^\mathrm{opt}=H_B\left(U_B^\mathrm{opt}\right).
$$

Differences between these functions and their quantum analogies are given by
\begin{eqnarray}
	 i_{A|B}^\mathrm{opt}-i_{A|B}=\frac{I_{AB}H_A^\mathrm{opt}-J_{AB}^\mathrm{opt}S_A}{S_AH_A^\mathrm{opt}}\geq0,\nonumber \\
	 i_{B|A}^\mathrm{opt}-i_{B|A}=\frac{I_{AB}H_B^\mathrm{opt}-J_{AB}^\mathrm{opt}S_B}{S_AH_B^\mathrm{opt}}\geq0, \nonumber
\end{eqnarray}
since we can use inequalities \cite{Renyi}
\begin{equation}\label{ineq}
	H_A^\mathrm{opt}\geq S_A, \quad H_B^\mathrm{opt}\geq S_B, \quad I_{AB}\geq J_{AB}^\mathrm{opt}.
\end{equation}

The situation with asymmetry, defined by
$$
	 d_{AB}^{\mathrm{opt}}=i_{A|B}^\mathrm{opt}-i_{B|A}^\mathrm{opt}=J^\mathrm{opt}_{AB}\frac{H_A^\mathrm{opt}-H_B^\mathrm{opt}}{H_A^\mathrm{opt}H_B^\mathrm{opt}},
$$
become quite nontrivial because the sign of difference $H_A^\mathrm{opt}-H_B^\mathrm{opt}$ can coincide or not with the sign of $S_A-S_B$. Therefore the conclusion about asymmetry between $A$ and $B$ made by original quantum causal analysis and in the considered ``optimal'' scheme can be quite different.

The relation between magnitudes
$$
	 \frac{d_{AB}^\mathrm{opt}}{d_{AB}}=\frac{J_{AB}^\mathrm{opt}}{I_{AB}}\frac{S_AS_B(H_A^\mathrm{opt}-H_B^\mathrm{opt})}{H_A^\mathrm{opt}H_B^\mathrm{opt}(S_A-S_B)}
$$
stays to be indefinite as well.

\subsection{``Tomographic'' scheme vs. ``optimal'' scheme}
The both approaches for studying observed classical correlations in quantum system seem to be quite natural, however, they are different.
The main feature of ``optimal'' (symmetric) discord is that it is greater or equal to ``tomographic'' one just because of maximization procedure
\begin{eqnarray}\label{eq:DiscIn}
	D^\mathrm{tom}_{AB}-D^\mathrm{opt}_{AB}= \qquad \qquad\qquad \\
	=\max_{U_A,U_B}{\{J_{AB}(U_A,U_B)\}}-J_{AB}(U_A^0,U_B^0)\geq0. \nonumber
\end{eqnarray}

On the other hand we can not say anything about difference between independence functions
\begin{eqnarray} i_{A|B}^\mathrm{opt}-i_{A|B}^\mathrm{tom}=\frac{J_{AB}^\mathrm{tom}H_A^\mathrm{opt}-J_{AB}^\mathrm{opt}S_A}{S_AH_A^\mathrm{opt}}\nonumber\\ i_{B|A}^\mathrm{opt}-i_{B|A}^\mathrm{tom}=\frac{J_{AB}^\mathrm{tom}H_B^\mathrm{opt}-J_{AB}^\mathrm{opt}S_B}{S_AH_B^\mathrm{opt}},\nonumber
\end{eqnarray}
since we can use (\ref{ineq}) and $J_{AB}^\mathrm{tom}\leq J_{AB}^\mathrm{opt}$.

The same is about relation between asymmetries
$$
 	 \frac{d_{AB}^\mathrm{opt}}{d_{AB}^\mathrm{tom}}=\frac{J_{AB}^\mathrm{opt}}{J_{AB}^\mathrm{tom}}\frac{S_AS_B(H_A^\mathrm{opt}-H_B^\mathrm{opt})}{H_A^\mathrm{opt}H_B^\mathrm{opt}(S_A-S_B)},
$$
which also depends on the particular quantum state.

We see the ``optimal'' scheme, which is interesting from the practical point of view as it affords maximal amount observed correlations, gives obscure conclusions about the original asymmetry, while ``tomographic'' scheme keeps original asymmetry properties but does not demonstrate maximal possible classical correlations. In the next section we are going to consider the both approaches to the two-qubit states which are the simplest examples of bipartite quantum systems.

\begin{figure*}
\center{\includegraphics[width=0.75\linewidth]{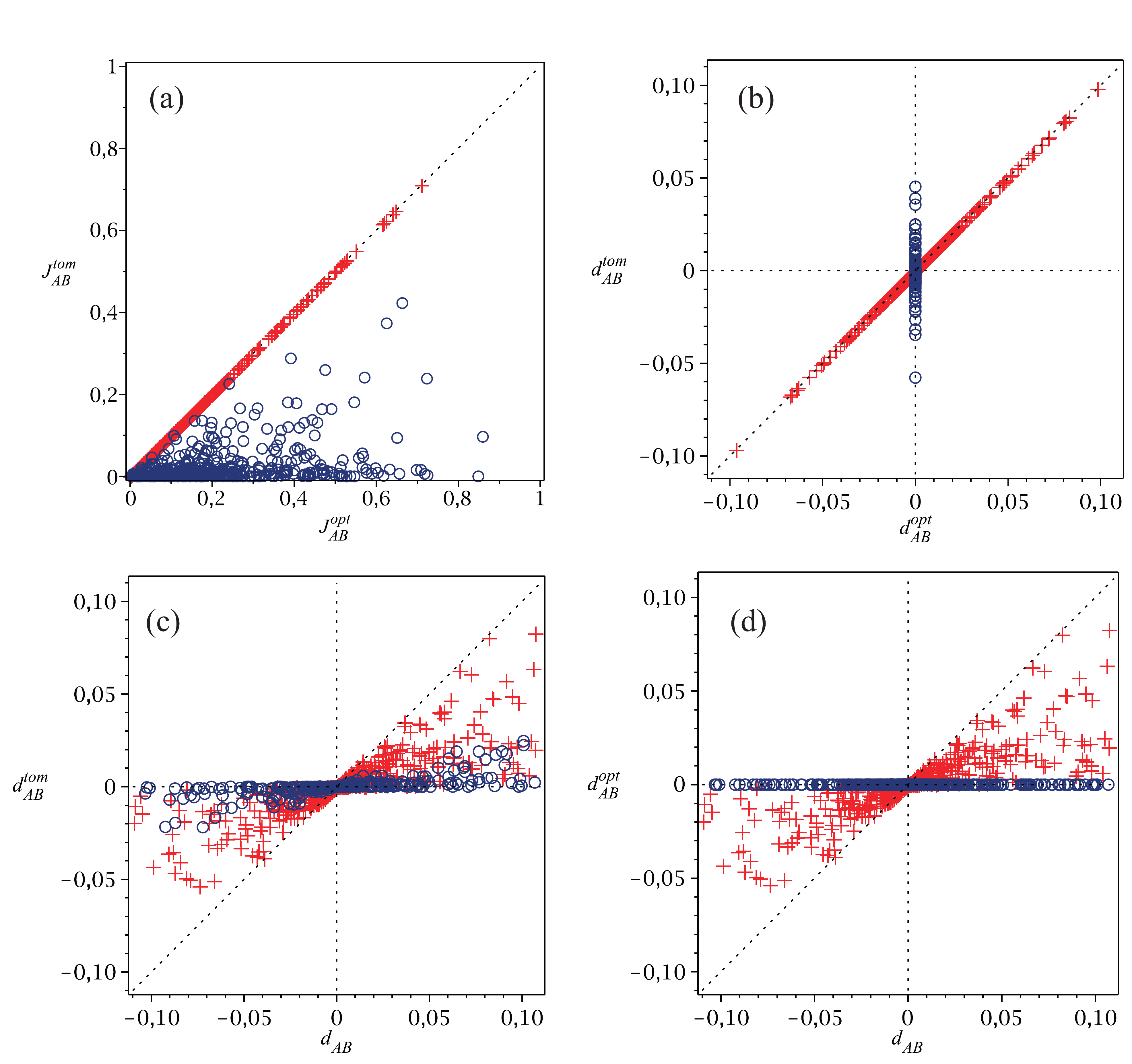}} \\
\caption{
Comparison of different characteristic for the data set of $N=1000$ randomly generated $X$-states:
(a) mutual information obtained in ``tomographic'' scheme {\it vs} the one in ``optimal'' scheme;
(b) asymmetry obtained in ``tomographic'' scheme {\it vs} the one in ``optimal'' scheme;
(c) asymmetry obtained in ``tomographic'' scheme {\it vs} the one obtained by quantum causal analysis;
(d) asymmetry obtained in ``optimal'' scheme {\it vs} the one obtained by quantum causal analysis. 
States with $\mathcal{T}_{AB}^\mathrm{tom}=\mathcal{T}_{AB}^\mathrm{opt}$ are depicted by crosses, all others -- by circles.}
\label{fig:XSt}
\end{figure*}

\section{Asymmetry in two-qubit states}\label{sec:5}
We are going to discuss three classes of two-qubit states.

(i) We start with the simplest case of pure two-qubit entangled states, which can be treated analytically.

(ii) Then we consider so-called $X$-states (states with definite constraints on density matrix).
They are generally mixed and we use method of random state generation for the purpose of their investigation (for a details, see Appendix I).

(iii) Finally, we look at arbitrary two-qubit mixed states, again by means of random state generation (Appendix II).

For these three classes we calculate all discussed measures of correlations ($I_{AB}$, $J_{AB}^\mathrm{tom}$, $J_{AB}^\mathrm{opt}$) and asymmetry ($d_{AB}$, $d_{AB}^\mathrm{tom}$, $d_{AB}^\mathrm{opt}$)
as well as investigate relation between them.

\subsection{Pure entangled states}
From the Schmidt decomposition it follows that any two-qubit pure entangled state $|\Psi\rangle_{AB}$ can be presented in the form
$$
	|\Psi\rangle_{AB}=\alpha|u_0,v_0\rangle_{AB}+\sqrt{1-\alpha^2}|u_1,v_1\rangle_{AB}, \quad \alpha\in(0,1),
$$
where sets
$$\{|u_0\rangle_A,|u_1\rangle_A\}, \quad \{|u_0\rangle_B,|u_1\rangle_B\}$$
produce bases in Hilbert spaces $\mathcal{H}_A$ and $\mathcal{H}_B$.

As it was already said, from the viewpoint of quantum causal analysis pure states are completely symmetric
$$
	i_{A|B}=i_{B|A}=-1, \quad d_{AB}=0.
$$

The ``tomographic'' and ``optimal'' approaches give the same conclusions about correlations
\begin{eqnarray}
	 J^\mathrm{tom}_{AB}=J_{AB}^\mathrm{opt}=-\alpha^2\log{\alpha^2}-(1-\alpha^2)\log{(1-\alpha^2)}, \nonumber
\end{eqnarray}
as well as about asymmetry
\begin{eqnarray}
	i_{A|B}^\mathrm{tom}=i_{B|A}^\mathrm{tom}=i_{A|B}^\mathrm{opt}=i_{B|A}^\mathrm{opt}=0, \quad d_{A,B}^\mathrm{tom}=d_{A,B}^\mathrm{opt}=0. \nonumber
\end{eqnarray}
Therefore, one can see that the alternative approaches to quantum causal analysis do not give any new results.

\begin{figure*}
\center{\includegraphics[width=0.75\linewidth]{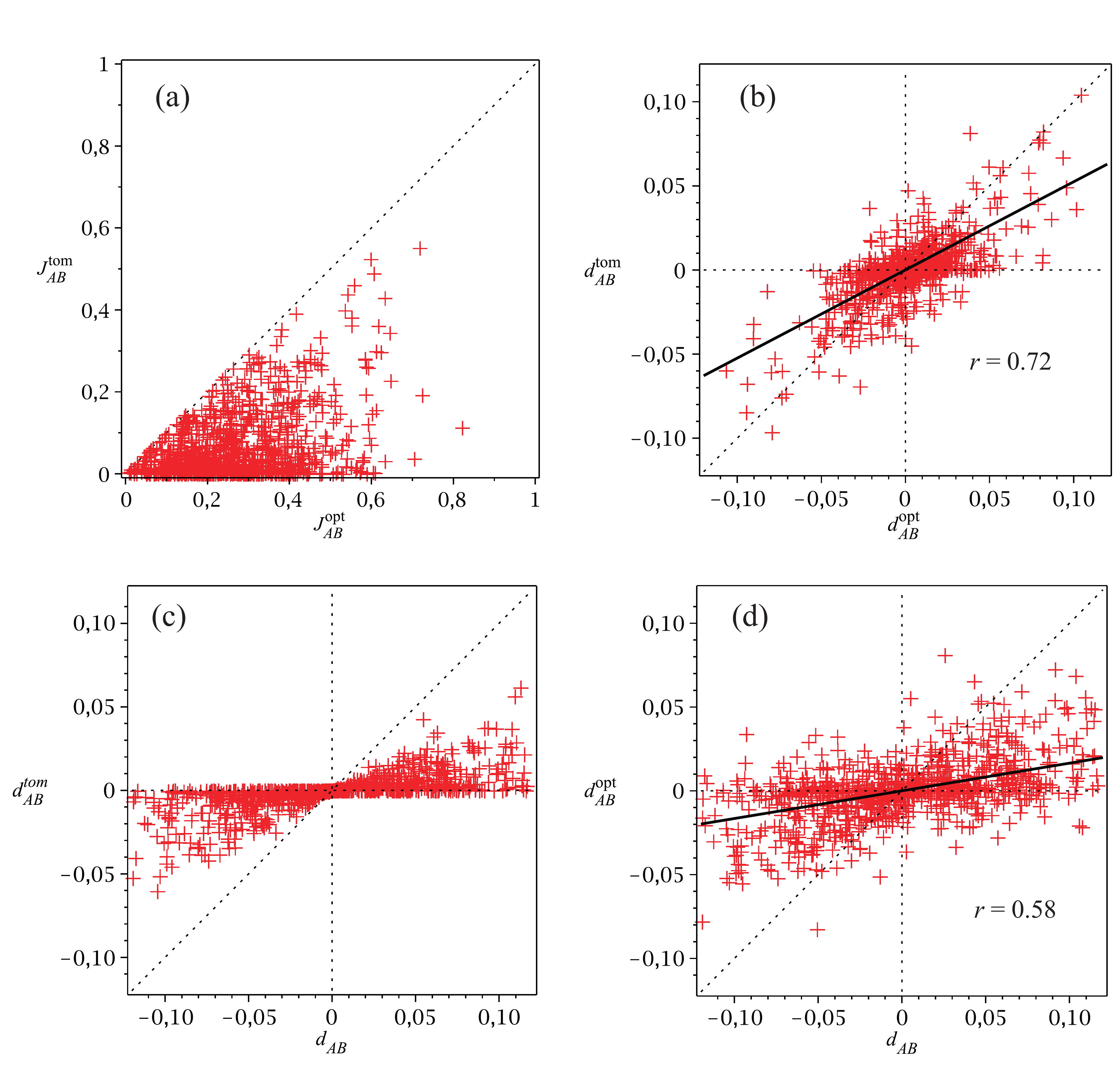}}
\caption{
Comparison of different characteristic for the data set of $N=1000$ randomly generated two-qubit states:
(a) mutual information obtained in ``tomographic'' scheme {\it vs} the one in ``optimal'' scheme;
(b) asymmetry obtained in ``tomographic'' scheme {\it vs} the one in ``optimal'' scheme.
(c) asymmetry obtained in ``tomographic'' scheme {\it vs} the one obtained by quantum causal analysis;
(d) asymmetry obtained in ``optimal'' scheme {\it vs} the one obtained by quantum causal analysis.
Solid lines stand for expressions of linear regression.}
\label{fig:RandSt}
\end{figure*}

\subsection{$X$-states}
The two-qubit $X$-state is the one that has a density matrix in the form of
\begin{equation}\label{eq:X-state}
\rho_{AB}=\begin{pmatrix}
	\rho_{11} & 0 & 0 & \rho_{14} \\
	0 & \rho_{22} & \rho_{23} & 0 \\
	0 & \rho_{23}^* & \rho_{33} & 0 \\
	\rho_{14}^* & 0 & 0 & \rho_{44} \\
	\end{pmatrix}.
\end{equation}

The constraints on matrix elements are the following:
(i) all the diagonal elements are positive $\left(\rho_{mm}\geq0\right)$;
(ii) trace is equal to unit: $\sum_{m=1}^{4}{\rho_{mm}}=1$;
(iii) $|\rho_{14}|^2\leq\rho_{11}\rho_{44}$, $|\rho_{23}|^2\leq\rho_{22}\rho_{33}$.
Its eigenvalues have the form
\begin{eqnarray}
	 \lambda_{1,2}=\frac{1}{2}\left(\rho_{11}+\rho_{44}\pm\sqrt{(\rho_{11}-\rho_{44})^2+4|\rho_{14}|^2}\right), \nonumber \\
	 \lambda_{3,4}=\frac{1}{2}\left(\rho_{22}+\rho_{33}\pm\sqrt{(\rho_{22}-\rho_{33})^2+4|\rho_{23}|^2}\right), \nonumber
\end{eqnarray}
that allows to obtain an explicit form of whole system von Neumann entropy: $S_{AB}=-\sum_{i=1}^4\lambda_i\log\lambda_i$.

The subsystem matrices are already diagonal
$$
\rho_{A}=\begin{pmatrix}
	\rho_{11}{+}\rho_{22} & 0 \\
	0 & \rho_{33}{+}\rho_{44}\\
	\end{pmatrix}, \,\
\rho_{B}=\begin{pmatrix}
	\rho_{11}{+}\rho_{33} & 0 \\
	0 & \rho_{22}{+}\rho_{44}\\
	\end{pmatrix},
$$
therefore $U_A^0=U_B^0=\mathrm{Id}_2$ are just $2\times2$ identity matrices.
This fact implies that
$$
	S_{A}=-\sum_{i=1}^{2}{ \rho_{A_{ii}}\log{\rho_{A_{ii}}}}, \,\, S_{B}=-\sum_{i=1}^{2}{ \rho_{B_{ii}}\log{\rho_{B_{ii}}}} \nonumber \\
$$

Due to the trivial form of $U_A^0$ and $U_B^0$ the tomogram $\mathcal{T}_{AB}^\mathrm{tom}$ just consists of diagonal elements: $\mathcal{T}_{AB}^\mathrm{tom}=\{\rho_{mm}\}_{m=1}^4$.
This allows to get straightforward expressions for all ``tomographic'' characteristics: $J_{AB}^\mathrm{tom}$, $i_{A|B}^\mathrm{tom}$, $i_{B|A}^\mathrm{tom}$ and $d_{AB}^\mathrm{tom}$.

Nevertheless, the question about optimal measurement bases, described by $U_A^\mathrm{opt}$ and $U_B^\mathrm{opt}$, keeps to be open.
In an effort to calculate $J_{AB}^\mathrm{opt}, i_{A|B}^\mathrm{opt}, i_{B|A}^\mathrm{opt}$ the iterative numerical algorithm has been used.

The main results for randomly generated data set of $N=1000$ $X$-states are presented in Fig.\ref{fig:XSt}.
The comparison of $J_{AB}^\mathrm{tom}$ and $J_{AB}^\mathrm{opt}$ reveals two types of $X$-states:

(i) for the first type (crosses in Fig.\ref{fig:XSt}) ``optimal'' basis turns to be the same as tomographic and the following equality holds
$$
	U_A^\mathrm{opt}=U_B^\mathrm{opt}=U_A^0=U_B^0={\rm Id_2}, \quad J_{AB}^\mathrm{tom}=J_{AB}^\mathrm{opt};
$$

(ii) for the second type (circles) the ``optimal'' basis is different from the ``tomographic'' and gives an advantage in correlations making the following inequality holds true
$$
	J_{AB}^\mathrm{opt}>J_{AB}^\mathrm{tom}.
$$

The detailed study of the states of second type has shown the ``optimal'' basis turns to be the one, which maximizes the both entropies: $H_A^\mathrm{opt}=H_B^\mathrm{opt}=1$, making the state to look symmetric with $d_{AB}^\mathrm{opt}=0$ (see Fig.\ref{fig:XSt}b).
This symmetrization implies the following form of unitary operators
$$
\begin{array}{c}
U_A^\mathrm{opt}=\begin{pmatrix}
	0 & e^{-i\phi_A} \\
	e^{i\phi_A} & 0\\
	\end{pmatrix} {\bf H}, \quad
U_B^\mathrm{opt}=\begin{pmatrix}
	0 & e^{-i\phi_B} \\
	e^{i\phi_B} & 0\\
	\end{pmatrix} {\bf H} \\
\end{array}
$$
with ${\bf H}$ being the $2\times2$ Hadamard transformation and $\phi_A$, $\phi_B$ being unique parameters for a particular $X$-state.

The comparison of $d_{AB}^{\mathrm{tom}}$ with $d_{AB}$, presented in Fig.~\ref{fig:XSt}c, shows a direct realization of the property~(\ref{eq:TomDeltaVSDelta}): the original asymmetry always turns to be stronger than the one obtained from $\mathcal{T}_{AB}(U_A^0\otimes U_B^0)$.

The relation between $d_{AB}^\mathrm{opt}$ and $d_{AB}$ (Fig.~\ref{fig:XSt}d) clearly demonstrates an existence of huge subclass of $X$-states with asymmetric quantum correlations and totally symmetric maximum available classical one.

\subsection{Arbitrary mixed states}

Finally, we are going to consider an arbitrary mixed two-qubit states in the general form
$$
	\rho_{AB}=\sum_{i=1}^{4}p_i|\Psi_i\rangle_{AB}\langle\Psi_i|, \quad \sum_{i=1}^{4}p_i=1
$$
with $|\Psi_i\rangle_{AB}$ being pure two-qubit state.

The obtained results for randomly generated data set of $N=1000$ arbitrary mixed states are presented in Fig.~\ref{fig:RandSt}.
The first plot (Fig.~\ref{fig:RandSt}a) is a demonstration of inequality (\ref{eq:DiscIn}): it shows that correlations in ``optimal'' scheme are always higher than in ``tomographic'' one.

The situation depicted in Fig.~\ref{fig:RandSt}b confirms the result from previous section that there is no explicit relation between $d_{AB}^\mathrm{tom}$ and  $d_{AB}^\mathrm{opt}$ for {\it arbitrary} state.
All combinations are possible: ``tomographic'' and ``optimal'' schemes can present the same or the opposite conclusions about a direction of asymmetry.
The strength of asymmetry in ``tomographic'' scheme also can be larger or smaller than the one in ``optimal'' scheme.

Nevertheless, Fig.~\ref{fig:RandSt}b shows an evident correlation between $d_{AB}^\mathrm{opt}$ and $d_{AB}^\mathrm{tom}$.
The exact value of correlation coefficient for the generated data set is $r=0.72$ and linear regression takes the form $d_{AB}^\mathrm{opt}=0.52d_{AB}^\mathrm{tom}$.
Taking it into account we can say that states with $d_{AB}^\mathrm{opt}d_{AB}^\mathrm{tom}>0$ are the more typical (in the generated data set $\approx70\%$ of states fulfills this condition).

The comparison of $d_{AB}^\mathrm{tom}$ with $d_{AB}$ in Fig.~\ref{fig:RandSt}c confirms inequality~(\ref{eq:TomDeltaVSDelta}): the direction of asymmetry given by ``tomographic'' scheme coincides with the one given by quantum causal analysis, while the magnitude of the original asymmetry is always higher than the observed one.

Finally, the comparison of $d_{AB}^\mathrm{opt}$ with $d_{AB}$ depicted in Fig.~\ref{fig:RandSt}d shows again the whole variety of possible combinations.
There is also a correlation between these two quantities but it is not so clear as in Fig.~\ref{fig:RandSt}b (correlation coefficient $r=0.58$ with linear regression $d_{AB}^\mathrm{opt}=0.17d_{AB}$).

\section{Conclusion}\label{sec:6}
\label{sec:Concl}

The original quantum causal analysis introduces a measure of asymmetry in bipartite quantum states based on inequality of their subsystems' von Neumann entropies.
To obtain this value one need to have a density operator of the studied system.
In the current Letter we have considered a novel approach to a definition of state's asymmetry, which is based on observable tomographic distributions of bipartite state.
As variation of measurement basis leads to a change of tomogram of state, we have restricted ourself with two variants of bases:
(i) the one that diagonalize a subsystems' density matrices,
(ii) the one that maximizes an observed amount of classical correlations between subsystems.
We have shown that these two approaches closely relate to tomographic and symmetric versions of quantum discord correspondingly, that is why we named the first approach ``tomographic'' and second one -- ``optimal''.

The comparison of ``tomographic'' modification of quantum causal analysis with the original one has shown that they give the same conclusion about direction of asymmetry, but the magnitude of asymmetry in ``tomographic'' scheme is always non-greater than the one obtained by the original method.

On the other hand, in the ``optimal'' scheme the both: direction and magnitude of asymmetry, can differ from the values obtained by original quantum analysis.

The implementation of these approaches to different classes of two-qubit states had shown the following main results:
(i) pure entangled states are always symmetric in respect to all approaches;
(ii) there are two subclasses of $X$-states: for the first one the ``tomographic'' and ``optimal'' approaches give the same results, and for the second one ``optimal'' approach demonstrates the full symmetry of the considered state, while original and ``tomographic'' approaches testifies some asymmetry presence;
(iii) in spite of the fact that ``tomographic'' and ``optimal'' schemes can give very different conclusion about asymmetry for the arbitrary two-qubit states there is an evident correlation between their results.

Finally, the question about practical aspect of obtained results appears.
As it was shown in previous works~\cite{Kikt2,Kikt3,Kikt4} the asymmetry of bipartite state observed by quantum causal analysis plays role in interaction of the such state with environment.
``Tomographic'' approach can reveal this asymmetry without full reconstruction of the state.
But questions about asymmetry ``optimal'' scheme keep to be open.
Is their any particularly features of states with different directions of ``optimal'' and ``tomographic'' (original) asymmetry?
Is their any protocols where the such features can play a crucial role?
All these questions seem to be important for a deeper understanding of how the results of local measurements performed on bipartite quantum system relates to its original properties as well as
in practical application in quantum information technologies.

\section*{Acknowledgements}

Authors thank V.I. Man'ko and S.N. Filippov for fruitful discussions.
We thank B.C. Sanders, S.M. Korotaev, and A.I. Lvovsky for useful comments.
A.K.F. was supported by RQC and Dynasty Foundation Fellowships.
E.O.K. was supported by Council for Grants of the President of the Russian Federation (grant SP-961.2013.5).
The work was supported by RFBR (12-05-00001 \& 14-08-00606).

\section*{Appendix I. Generation of $X$-states} \label{sec:Ap1}
For the generation of $X$-states (\ref{eq:X-state}) the following algorithms was used.
The diagonal elements were generated in the form
$$
	\rho_{ii}=\frac{p_i}{\sum_{j=1}^{4}p_j}, \quad p_i=\mathcal{U}(0,1),
$$
where $\mathcal{U}(a,b)$ stands for uniform distribution in $[a,b]$.
Non-diagonal elements were generated in the form
\begin{eqnarray}
	\rho_{14}=\rho_{41}^*=\alpha_{1}\sqrt{\rho_{11}\rho_{44}}\,e^{i\phi_1}, \quad \alpha_{1(2)}=\mathcal{U}(0,1) \nonumber \\
	\rho_{23}=\rho_{32}^*=\alpha_{2}\sqrt{\rho_{22}\rho_{33}}\,e^{i\phi_2}, \quad \phi_{1(2)}=\mathcal{U}(0,2\pi) \nonumber
\end{eqnarray}
In spite of the fact that considered method does not generate states in uniform way in respect to the Haar measure it is quite useful for the purposes of our study.

\section*{Appendix II. Generation of arbitrary two-qubit mixed state}\label{sec:Ap2}
The arbitrary two-qubit mixed states were generated in the form
$$
	\rho_{AB}=\frac{1}{\sum_{j=1}^{4}p_j} \sum_{k=1}^4 p_k \langle\psi_k|\psi_k \rangle^{-1} |\psi_k\rangle_{AB}\langle\psi_k|,
$$
where
$$
	|\psi_k\rangle=\begin{pmatrix}
	\mathcal{N}(0,1) \\
	\mathcal{N}(0,1) \\
	\mathcal{N}(0,1) \\
	\mathcal{N}(0,1)
	\end{pmatrix}+i\begin{pmatrix}
	\mathcal{N}(0,1) \\
	\mathcal{N}(0,1) \\
	\mathcal{N}(0,1) \\
	\mathcal{N}(0,1)
	\end{pmatrix},
$$
with $p_k=\mathcal{U}(0,1)$ and $\mathcal{N}(\mu,\sigma)$ being normal distribution with expectation $\mu$ and standard deviation $\sigma$.
According to Ref.~\cite{StateGen} this method gives uniform distribution of states.


\begin{thebibliography}{10}
\bibitem{disc}
The question about connection between asymmetry in bipartite states and real causality is interesting, however, it is beyond the present work scope.
\end{thebibliography}
\end{document}